\documentclass{elsart}  

\usepackage{amsmath}  
\usepackage{epsfig}   

\renewcommand{\bar}[1]{\overline{ #1}} 
\newcommand{\D}{\mathrm{d}} 
\newcommand{\I}{\mathrm{i}} 

\begin{document}


\begin{frontmatter}  
    
  
    \title{Regularization for effective field theory with two heavy particles\thanksref{Preprint}}
    
    \author[HU,JLab]{J.~L.~Goity,}
    \author[HU,JLab]{D.~Lehmann,}
    \author[HU]{G.~Pr{\'e}zeau\thanksref{Mailaddress},}
    \author[HU]{and J.~Saez}
    
    \address[HU]{Department~of~Physics, Hampton~University, Hampton, VA~23668, USA}
    \address[JLab]{Thomas~Jefferson~National~Accelerator~Facility,
      12000~Jefferson~Avenue~M.S.~12H2, Newport~News, VA~23606, USA}
    \thanks[Preprint]{\texttt{JLAB-THY-00-46}}
    \thanks[Mailaddress]{Mail address: Thomas~Jefferson~National~Accelerator~Facility,
      12000~Jefferson~Avenue~M.S.~12H2, Newport~News, VA~23606, USA}
    

    \begin{abstract}
      A regularization for effective field theory with two propagating
      heavy particles is constructed. This regularization preserves
      the low-energy analytic structure,  implements a low-energy power
      counting for the one-loop diagrams, and preserves symmetries respected
      by dimensional regularization.
    \end{abstract}
    
    \begin{keyword}
      Regularization \sep
      Effective Field Theory \sep
      Baryon Chiral Perturbation Theory \sep
   NN Effective Field Theory \sep
   Low-Energy Power Counting.
      \PACS 11.10.-z \sep 11.15.Bt \sep 12.39.Hg \sep 12.39.Fe 
    \end{keyword}

  \end{frontmatter}


\section{Introduction}
\label{sec:Introduction}

One of the necessary  ingredients in an effective field theory (EFT) is the
ultraviolet regularization. A natural regularization for the EFT is one that
removes  from loop integrals all contributions with momenta above the 
EFT energy scale, $\Lambda_{\text{EFT}}$, and that also implements a low-energy
 counting by making the small energy scales in the EFT, namely small momenta
and  light particle masses, the scales that set the magnitude of  power
ultraviolet divergences. One example of such a natural
regularization is dimensional regularization in EFT with only light particles,
such as Chiral Perturbation Theory (ChPT)  \cite{Weinberg1,GL1} for
quasi-Goldstone pions. In dimensional regularization the scale of ChPT is
given in terms of the pion decay constant $F_\pi$, and power divergences of
loop integrals are proportional to the small scales ($M_\pi^2$ and/or 
invariants formed with the small external momenta) divided by the appropriate
powers of $F_\pi$. 
Although a momentum cutoff can be used, it does  not deal well with power
divergences, as these contain pieces proportional to powers of the cutoff.  
In this case,  an EFT can also be formulated, but it is more complicated to
deal with because  the low-energy power counting must be recovered in the
process of renormalization. 

In an EFT with a heavy stable particle, such as ChPT with nucleons,
the heavy particle mass $M$ is a new scale in the theory (some heavy masses
could be much larger than $\Lambda_{\text{EFT}}$, e.g., baryon masses in large-$N_{\text{c}}$
QCD). With the inclusion of $M$,  dimensional regularization ceases to be a natural regularization. 
Indeed, the power divergences lead now to contributions proportional to
fractional powers of $M$, i.e.\  contributions that give rise to terms that are
powers of $M^2$ times  $\log M^{2}$ in $d\to 4$ dimensions. One such example is
the nucleon self-energy in ChPT calculated in dimensional regularization
\cite{GSS}.  Although the EFT can be consistently regularized with dimensional
regularization, a natural regularization is more convenient.

In the case, where a single heavy particle propagates through a Feynman diagram, 
one way to remove  fractional powers of $M$ \cite{BL} is to perform an
expansion in $1/M$ at the level of the Lagrangian \cite{JM}, eliminating in
this way $M$ from the propagators. This approach implies  that all the Green
functions must be analytic functions of $1/M$ in the low-energy region.
This, however, is not correct because there are low-energy singularities in 
$1/M$, such as the  anomalous thresholds (e.g., in the scalar form factor of the 
nucleon \cite{M}),  which the EFT must retain.
Recently, a regularization based on dimensional regularization
that avoids these problems of the $1/M$ expansion was developed.
Building on work by  Ellis and Tang \cite{ET},  
Becher and Leutwyler \cite{BL} developed a regularization scheme that is natural,  
preserves the low-energy structure and internal symmetries, and is Lorentz
covariant. This regularization has been called infrared regularization.

In the case of two heavy particles,  the $1/M$
expansion cannot be applied due to the two-heavy particle threshold in the
s-channel. It is, therefore, very important to develop a natural
regularization for this case as well. In this letter such a regularization  is
developed along lines similar to those of the infrared regularization
\cite{BL}. Starting from standard dimensional regularization, a modification
of the Feynman  parameter integration domain is introduced leading to a
natural regularization. Its  implementation is presented in a Lorentz
covariant framework for basic one-loop diagrams.

\section{Regularization of one-loop diagrams}
\label{sec:OneLoop}

In this section, the basic one-loop diagrams in the two-heavy-particle sector are
discussed: these are the bubble diagram Fig.~\ref{fig:Bubble}, 
the triangle diagram Fig.~\ref{fig:Triangle} and the box diagram Fig.~\ref{fig:Box}.  
In each case, the dimensionally regularized Feynman integral  $I$ can be 
decomposed into
\begin{equation}
  \label{eq:Decomp}
  I = \bar{I} + R\:,
\end{equation}
where the fractional powers of $M$ are entirely contained in $R$.
In addition, $R$ is an analytic function in the low-energy domain
(i.e. local) and is referred to as the regular part. The low-energy part
$\bar{I}$ contains all the low-energy analytic structure of the integral.
Because $\bar{I}$ does not contain any fractional powers of $M$, it satisfies
a low-energy power counting as discussed in Sect.~\ref{sec:Discussion}.
The loop integral in the EFT
Dimensional Regularization (EFTDR) is then identified with $\bar{I}$.

The low-energy domain is expressed in terms of the heavy mass $M$, the light mass $m$,
and the invariants $P_{i}\cdot P_{j}$ formed with the external heavy particle momenta, 
and is defined by the conditions $\mid\! P_{i}\cdot P_{j} - M^{2} \!\mid \ll \Lambda_{\text{EFT}}^{2}$
and $m \ll \Lambda_{\text{EFT}}$. As in ChPT, low energy quantities such as
$m$, external momenta of light particles, differences of external momenta of
heavy particles,  etc., will be said to be of order $p$.

Although integrals that carry factors of the
loop momentum in the numerator are not explicitly presented here, such integrals 
can be reduced to combinations of scalar integrals of the form discussed here, 
to integrals that involve only one heavy propagator that can be regularized
using  the  infrared regularization, and to integrals that involve only light
propagators.

\subsection{The bubble diagram}
\label{sec:Bubble}

\begin{figure}[htbp]
  \centerline{\epsfysize=3.5cm \epsfbox{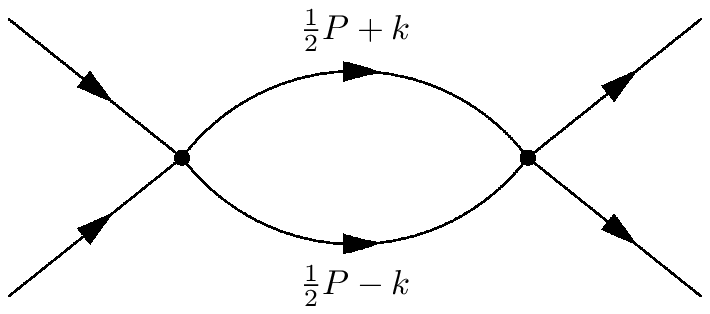}} 
  \caption{}
  \label{fig:Bubble}
\end{figure}

The simplest diagram in the two-heavy-particle sector is the bubble diagram 
of Fig.~\ref{fig:Bubble} with  the two heavy particles propagating in the
s-channel.   The scalar integral for this diagram is
\begin{eqnarray}
  I_2(s,M^2)&=&\mu^{4-d}\!\int\!\!\frac{\mbox{d}^dk}{(2\pi)^d} \: \frac{\mbox{i}}{
    (\frac{P}{2}-k)^2-M^2+\mbox{i}\epsilon}\;
  \frac{\mbox{i}}{(\frac{P}{2}+k)^2-M^2+\mbox{i}\epsilon} \:,
\end{eqnarray}
where $s \equiv P^{2}$ and $P$ is the total incoming momentum.
Combining the propagators using a Feynman parameter integral, the
integration over momentum yields
\begin{eqnarray}\label{bubble}
  I_2=\kappa_2 \int^1_{-1}\!\!\mbox{d}w \:
  \bigl[\frac{s}{4}(w^2-1) + M^2  - \I \epsilon\bigr]^{d/2-2}\:,
\end{eqnarray}
with
\begin{eqnarray}\label{kappa}
  \kappa_2\equiv -\frac{\mbox{i}}{2} \frac{1}{(4 \pi)^{2}} \Gamma(2-\frac{d}{2}) (4 \pi \mu^{2})^{2-d/2}
  \:.
\end{eqnarray}
This integral has a pinch singularity at $w=0$ 
when $s = s_{+} \equiv 4M^2$, which is the branch point of the two-heavy-particle cut.
The integration boundaries give rise to fractional powers of $M$ which spoil low-energy
power counting. This can be shown below threshold by scaling the small quantity
$(M^2 - s/4)$ out of Eq.~(\ref{bubble}):
\begin{eqnarray}\label{scaledbubble}
  I_2=\kappa_2\left(\frac{s}{4}\right)^{d/2-2} \alpha^{d-3}
  \int^{\frac{1}{\alpha}}_{-\frac{1}{\alpha}}\!\!\mbox{d}w\: 
  (w^2 +1)^{d/2-2}\:,
\end{eqnarray}
where
\begin{eqnarray}\label{alphan}
\alpha\equiv \sqrt{\frac{s_{+}-s-\I \epsilon}{s}}
\end{eqnarray}
is a small quantity in the low-energy domain which characterizes the infrared behavior
of the bubble diagram.
From Eq.~\eqref{scaledbubble} it is clear that only the integral limits can change the
power of $\alpha$.

The integral can then be separated into its regular and low-energy part: 
\begin{eqnarray}
  I_2=\bar{I}_{2}+R_{2}\:,
\end{eqnarray}
with
\begin{eqnarray}\label{Ibar}
  \bar{I}_{2}\equiv\kappa_2
  \left(\frac{s}{4}\right)^{d/2-2} \alpha^{d-3}
  \int^\infty_{-\infty}\!\!\mbox{d}w \: (w^2 +1)^{d/2-2}\:.
\end{eqnarray}
$\bar{I}_2$ and $I_2$ have the same low-energy analytic structure as
can be verified for $d\to 4$, where
\begin{eqnarray}\label{remainingI}
  I_2 &=&
\frac{\mbox{i}}{16\pi^2}\left[-\Gamma(2-\frac{d}{2})+
    \log\left(\frac{M^2}{4\pi\mu^2}\right) -2 +
    2\alpha\arctan\left(\frac{1}{\alpha}\right) \right]\:,
\end{eqnarray}
and where, after performing one integration by parts,
\begin{eqnarray}\label{IbarR}
  \bar{I}_2=\frac{\mbox{i}}{16\pi}\alpha\:.
\end{eqnarray}
  $\bar{I}_2$  is, therefore, given by the term of order $\alpha$ in the Taylor
expansion of Eq.~(\ref{remainingI}) in powers of $\alpha$, as  can be
immediately seen from  
\begin{eqnarray}
  \alpha \arctan\left(\frac{1}{\alpha}\right)&=
  \frac{\pi}{2}\alpha - \alpha
  \arctan\alpha,
\end{eqnarray}
The term $\alpha\arctan\alpha$ is analytic in $s_+-s$ in the
low-energy region. It is then clear that $\bar{I}_2$ and $I_2$ have the same
low energy analytic structure.

In arbitrary d-dimensions, $\bar{I}_2$  is proportional to
$(s_+-s)^{(d-3)/2}/\sqrt{s}$, which implies that it does not contain fractional
powers of $M$.   On the other hand, for $d\to 4$, $R_2$ is given by
 \begin{eqnarray}  
R_2=2\frac{M^{d-4}}{d-3}\kappa_2-\frac{\mbox{i}}{8\pi^2}\alpha  
\arctan\alpha ,
\end{eqnarray}
  while in the low-energy domain
and for arbitrary $d$
\begin{eqnarray}
  R_2&=&2\kappa_2M^{d-4}\sum\limits_{m,n=0}^\infty c_{m,n}\frac{(s_+
-s)^{m+n}}{s_+^m s^n},
 \end{eqnarray}
where
\begin{eqnarray}
c_{m,n}\equiv \frac{(-1)^m
\Gamma^2\!(\frac{d}{2}-1)}{(d-3-2n)m!n!\Gamma(\frac{d}{2}-1-m)
\Gamma(\frac{d}{2}-1-n) },
\end{eqnarray}
which shows that $R_2$ is given by terms analytic in
$s-s_+$ multiplied by a fractional power of  $M$.
In particular, this shows that the analytic terms not included in $\bar{I}_2$
are all of fractional-power type.

The separation into a low-energy part and a regular part with the desired
properties has been achieved. 
Indeed, from Eqs.~(\ref{alphan}) and
(\ref{IbarR})  it is seen that $\bar{I}_2$ has two cuts: a cut due to the
factor $\sqrt{s_{+}-s}$ starting  at $s=s_{+}$ and a cut along the negative
axis ending at  $s=0$ due to the factor of  $1/\sqrt{s}$.  
The discontinuity of $I_{2}$ and $\bar{I}_{2}$ across the 
cut on the positive $s$-axis are equal, and therefore  $\bar{I}_2$
reproduces the low energy analytic structure of $I_2$. 
The cut
along the negative $s$-axis is located far outside the low-energy region.  Moreover,
$\bar{I}_{2}$ does not carry any fractional power  of $M$.
On the other hand, $R_2$ is purely of fractional power type,  and has only the
cut along the negative $s$-axis.

\subsection{The triangle diagram}
\label{sec:Triangle}

\begin{figure}[htbp]
  \centerline{\epsfysize=4.5cm  \epsfbox{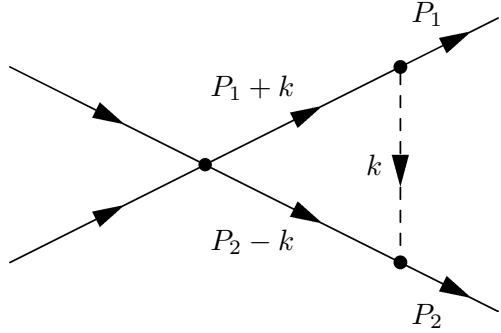}} 
  \caption{Triangle diagram in the two heavy-particle-sector. 
    The solid line represents the heavy particle, the dashed line the light particle.}
  \label{fig:Triangle}
\end{figure}

The next diagram to be considered  is the triangle
diagram shown in   Fig.~\ref{fig:Triangle}
with  two heavy particles propagating in the $s$-channel 
and interacting via a contact interaction and the exchange of a light
particle of mass $m$. 

The scalar integral for this diagram is
\begin{multline}
  \label{eq:DefTriangle}
  I_{3}(s,P_{1}^{2},P_{2}^{2},m^2,M^2)  \\
  = \mu^{4-d} \:\int\!\!\frac{\D^{d}k}{(2\pi)^{d}}\: 
  \frac{\I}{(P_{1}+k)^2-M^{2}+\I \epsilon}\,  
  \frac{\I}{(P_{2}-k)^2-M^{2}+\I \epsilon}\,
  \frac{\I}{k^2-m^{2}+\I \epsilon} 
  \:.
\end{multline}
Combining the heavy propagators with a symmetric Feynman parameter integral,
\begin{equation}
  \label{eq:SymFeynParam}
  \frac{1}{A_{1} \, A_{2}} = \frac{1}{2} \int_{-1}^{1} \frac{\D w}{\left(\frac{
1+w}{2} A_{1} + \frac{1-w}{2} A_{2} \right)^{2}}\:, \end{equation}
and subsequently combining this result with the light propagator using
\begin{equation}
  \label{eq:PiNFeynParam}
  \frac{1}{a \, A^{2}} = 2 \int_{0}^{1} \frac{z \, \D z}{\left((1-z) a + z A 
\right)^{3}}\:, \end{equation}
the momentum integration yields
\begin{equation}
  \label{eq:FSI}
  \begin{split}
    I_{3} &= \kappa_{3} 
    \int_{0}^{1}\!\!\D z \int_{-1}^{1}\!\!\D w\: 
    z \bigl[ C(w,z) - \I \epsilon \bigr]^{d/2-3} \:, 
    \\
    \kappa_{3} &= - \frac{1}{16 \pi^2} \:  \frac{1}{2} \: 
     \Gamma(3-\frac{d}{2}) \: (4\pi \mu^{2})^{2-d/2}  \:,
    \\
    C(w,z) &= (1-z) m^2 + z M^2 - z^2 (1-w^2) \frac{s}{4}  \\
    &\qquad - z(1-z) \frac{1}{2} (P_{1}^{2}+P_{2}^{2}) 
    - w z(1-z) \frac{1}{2} (P_{1}^{2}-P_{2}^{2})  \:. 
  \end{split}
\end{equation}

\begin{figure}[htbp]
  \centerline{\epsfysize=6cm \epsfbox{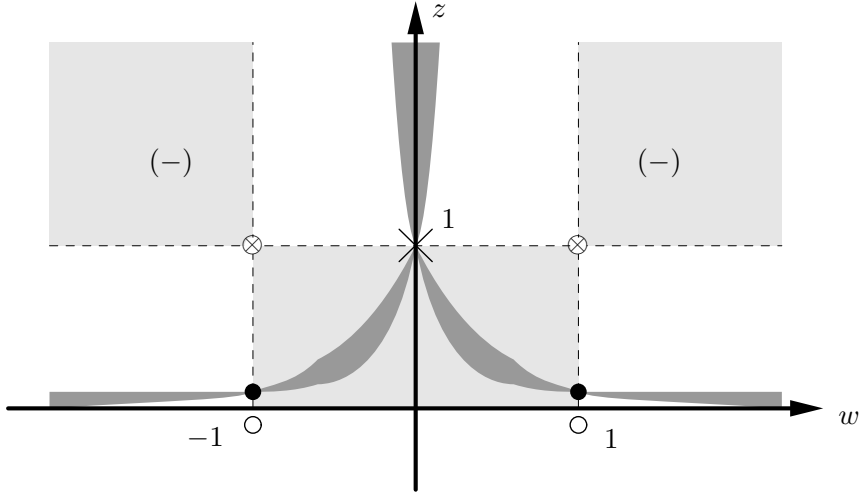}} 
\vspace{8mm}  
\caption{Triangle diagram: Integration domain and location of singularities
in the $(w,z)$-plane.      The light shaded domains indicate the contour of
integration of the regularized part $\bar{I}_{3}$,      areas marked with
$(-)$ enter with negative sign. The dark shaded area denotes critical points  
  associated with the leading singularity of the triangle diagram. The
two-heavy-particle ($\times$) and the     two heavy-light ($\bullet$)
thresholds lie on the boundary of the original integration domain
$[-1,1]\times[0,1]$.      Also indicated are the heavy-light pseudo-thresholds
($\circ$). The corner singularities ($\otimes$) give     rise to fractional
powers of $M$.}   \label{fig:LandauPlot} \end{figure}

In order to decompose $I_{3}$ into its low-energy  and regular parts,
\begin{equation}
  \label{eq:Decomposition}
  I_{3}= \bar{I}_{3} + R_{3}\:,
\end{equation}
the integration range in $w$ is first extended to the entire real axis in analogy to the
bubble diagram,
\begin{equation}
  \label{eq:ContributionA}
  \bar{I}_{3 \text{A}} = \kappa_{3} 
    \int_{0}^{1}\!\!\D z \int_{-\infty}^{\infty}\!\!\D w \: 
    z \bigl[ C(w,z) - \I \epsilon \bigr]^{d/2-3} \:. 
\end{equation}
$\bar{I}_{3 \text{A}}$ contains the proper singularity structure associated with the two-heavy-particle 
threshold at $s=s_{+}$ but does not contain the two-particle cuts corresponding to the  
heavy-light thresholds at $P_{i}^{2} = (M+m)^{2}$, ($i=1,2$).
All this becomes clear by solving the corresponding Landau equations\footnote{%
  For a review on Landau equations see Refs.~\cite{Eden,Itzykson,BD} and
references therein.} that determine the location of the singularities in the
$(w,z)$-parameter plane shown in Fig.~\ref{fig:LandauPlot}. The two-particle
thresholds are endpoint singularities, i.e., they lie on the boundary of the
original integration domain $[-1,1]\times [0,1]$.  The two-heavy-particle cut
(indicated by $\times$ in Fig.~\ref{fig:LandauPlot}) occurs at $(0,1)$ and  is
unaffected by extending the domain as in Eq.~\eqref{eq:ContributionA}. On the
other hand, the heavy-light thresholds (indicated by $\bullet$) are located at
 $(+1,\frac{m}{M+m})$ and  $(-1,\frac{m}{M+m})$,  and are lost.
The heavy-light thresholds, indeed the full low-energy analytic structure, are restored by
defining the low-energy part as
\begin{equation}
  \label{eq:TriangleIRPart}
  \bar{I}_{3} = \bar{I}_{3 \text{A}} + \bar{I}_{3 \text{B}}\:,
\end{equation}
with 
\begin{equation}
  \label{eq:ContributionB}
  \bar{I}_{3 \text{B}} = - \kappa_{3} 
  \int_{0}^{\infty}\!\!\D z  \left(\int_{-\infty}^{-1}\!\!\D w + \int_{1}^{\infty}\!\!\D w\right)
  \: z \bigl[ C(w,z) - \I \epsilon \bigr]^{d/2-3} \:.
\end{equation}
It is evident from Fig.~\ref{fig:LandauPlot} that $\bar{I}_{3}$ properly 
incorporates all two-particle thresholds at low energies 
and does not pick up any new low-energy endpoint singularities on the 
physical sheet.\footnote{In particular, the pseudo-thresholds 
  $P_{i}^{2} = (M-m)^2$ located at $(\pm 1, -m/(M-m))$ and indicated by $\circ$ in
  Fig.~\ref{fig:LandauPlot} are \emph{not} promoted to thresholds on the physical sheet.}
It is also important to note that $\bar{I}_{3}$ receives no additional contribution from the 
leading (i.e. pinch) singularity (dark shaded area in Fig.~\ref{fig:LandauPlot}) 
other than that already present in $I_{3}$. 
This ensures that the analytic properties of $\bar{I}_{3}$ and $I_{3}$ coincide
in the low-energy domain and implies that $R_{3}$ is analytic there.

By virtue of Eqs.~\eqref{eq:Decomposition} and \eqref{eq:TriangleIRPart} the regular part is
\begin{align}
  R_{3} &= \kappa_{3} 
  \int_{1}^{\infty}\!\!\D z  \left(\int_{-\infty}^{-1}\!\!\D w + \int_{1}^{\infty}\!\!\D w\right)
  \: z \bigl[ C(w,z) - \I \epsilon \bigr]^{d/2-3} \:. \label{eq:TriRegPart}
\end{align}
The dependence on fractional powers of $M$ stems from the corner points located at $(\pm 1,1)$
(indicated by $\otimes$ in Fig.~\ref{fig:LandauPlot}) and is entirely contained in
$R_{3}$.
It is instructive to verify this dependence on fractional powers of $M$ explicitly.
The denominator in Eq.~\eqref{eq:TriRegPart} is 
non-vanishing and of zeroth order in the low energy expansion in the whole
integration domain and can, therefore, be expanded in $m^{2}$ and the small 
variables $s_{+}-s$, $P_{i}^{2}-M^{2}$: \begin{equation}
   C(w,z) =  M^{2} z^{2} w^{2} + \mathcal{O}(s_{+}-s; P_{i}^{2}-M^{2}; m^{2})\:.
\end{equation}
The leading contribution in an expansion of $R_{3}$ is then
\begin{align}
  R_{3} &= 2 \kappa_{3} M^{d-6} \frac{1}{d-4} \: \frac{1}{d-5} \: \left(1+\cdots \right) 
  \label{eq:FracMass}\\
  &= - \frac{1}{16 \pi^{2}} \frac{1}{2 M^{2}} 
  \left( \frac{2}{4-d} -2 -\gamma - \log \frac{M^{2}}{ 4 \pi \mu^{2}} \right)
  \: \left(1 + \cdots \right)\:,\label{eq:FracMass4d}
\end{align}
where the omitted terms are analytic in the small quantities. The reason why
$R_3$ has an ultraviolet divergence despite $I_3$ being ultraviolet finite is
that in order to contain the fractional powers of $M$ while being regular in
the low-energy domain, it must include the appropriate logarithmic dependence
on $\mu$, and this can only emerge if there is a $1/(d-4)$ singularity.

The low-energy parts are shown below for $d\to 4$. The integral $\bar{I}_{3A}$
is given by:
 \begin{equation}
\label{eq:IBAR3A}
\bar{I}_{3 \text{A}} =-\frac{1}{16 \pi \sqrt{s}} \int_0^1{\rm
d}z\frac{1}{\sqrt{C_0}}\; , 
\end{equation}
with
\begin{equation}
\begin{split}
C_0&=z M^2+(1-z) m^2-\frac{z^2 s}{4}-\frac{1}{2}z(1-z)
(P_1^2+P_2^2)-\frac{(1-z)^2}{4 s}(P_1^2-P_2^2)^2 -\I\epsilon\:.
\end{split}
\end{equation}

and the integral $\bar{I}_{3B}$ is on the other hand given by:
\begin{multline}
  \bar{I}_{3\text{B}} = \frac{1}{16\pi^2} \biggl\{
  \left(\frac{2}{4-d} -\gamma- \log \frac{m^{2}}{4 \pi \mu^{2}} \right)
  \left(\frac{2}{s} + \cdots\right)\\
  + \frac{1}{2}  \left( \int_{-\infty}^{-1}\!\!\D w + \int_{1}^{\infty}\!\!\D w\right)
  \frac{z_{0}}{\sqrt{D_0 \;D_1}} 
\left( \frac{\pi}{2}+
\arctan\left(\sqrt{\frac{D_1}{D_0}} z_0 \right)\right)\biggr\}\:,
\end{multline} 
with
\begin{equation}
  \begin{split}
    z_{0} &= \frac{1}{4 D_{1}} \left( 2 (m^{2}-M^{2}) + (1+w) P_{1}^{2} + (1-w)
P_{2}^{2} \right) \:,\\     D_{1} &= \frac{1}{2} \left( (1+w) P_{1}^{2} + (1-w)
P_{2}^{2} -  (1-w^{2}) \frac{s}{2} \right)\:,\\   
   D_{0} &= m^2 - \frac{1}{4
D_{1}} \left(     M^2-m^2- \frac{1}{2} (1+w) P_{1}^{2} - \frac{1}{2} (1-w)
P_{2}^{2} \right)^2-\I\epsilon \:.   \end{split}
\end{equation}
It is to be noted that the integral $\bar{I}_{3A}$
has no ultraviolet divergences. Such
divergences can only come from the boundary contributions in the integration
by parts over $w$, which in this case have to vanish because the domain of
integration extends to the entire real axis. The low energy parts do not contain any fractional powers
of $M$, as their explicit evaluation reveals the absence of $\log{M^2}$ terms.
Thus, as in the case of the bubble diagram, the separation into low-energy and
regular parts is also possible in the triangle diagram. The general procedure
is now clear and can be  easily carried over to other diagrams. As an
illustration, we show this for the box diagram.

\subsection{The box diagram}
\label{sec:Box}

\begin{figure}[htbp]
  \centerline{\epsfysize=5.5cm \epsfbox{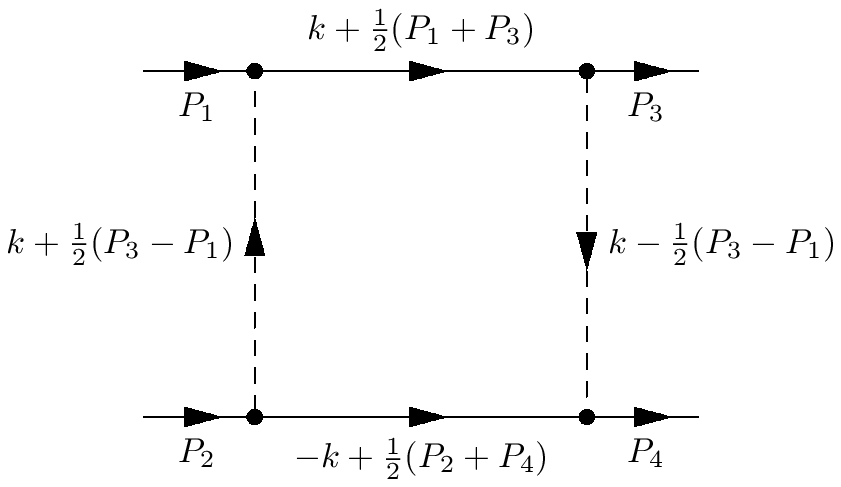}}
  \caption{}
  \label{fig:Box}
\end{figure}

The scalar integral for the box diagram in Fig.~\ref{fig:Box} is
\begin{eqnarray}\label{box}
  I_4&=&\mu^{4-d}\!\!\int\!\!\frac{\mbox{d}^dk}{(2\pi)^d}\:
  \frac{\mbox{i}}{(\frac{P_1+P_3}{2}+k)^2\!-\!M^2\!+\!\mbox{i}\epsilon} 
 \; \frac{\mbox{i}}{(\frac{P_2+P_4}{2} -k)^2\!-\!M^2\!+\!\mbox{i}\epsilon}
  \nonumber\\
  & &\hspace{7mm}
  \times\frac{\mbox{i}}{(\frac{P_3-P_1}{2}
    -k)^2\!-\!m^2\!+\!\mbox{i}\epsilon} 
 \; \frac{\mbox{i}}{(\frac{P_3-P_1}{2}
    +k)^2\!-\!m^2\!+\!\mbox{i}\epsilon}.
\end{eqnarray}
The light particle propagators can be combined to yield:
\begin{eqnarray}\label{Cbox}
  I_4=\mu^{4-d}\!\frac{\partial}{\partial m^2}
  &\int_0^1&\mbox{d}x\int\!\frac{\mbox{d}^dk}{(2\pi)^d}\;
  \frac{1}{[k-\bar{q}(x)]^2-\bar{m}^2(x)+\!\mbox{i}\epsilon}
  \nonumber\\ &\times&
  \frac{\mbox{1}}{(\frac{P_1+P_3}{2}
    +k)^2\!-\!M^2\!+\!\mbox{i}\epsilon} 
  \frac{\mbox{1}}{(\frac{P_2+P_4}{2}
    -\!k)^2\!-\!M^2\!+\!\mbox{i}\epsilon},
\end{eqnarray}
with
\begin{displaymath}
  \bar{q}(x)\equiv \frac{1}{2}(P_3-P_1)(1-2x) \qquad \text{and}
  \qquad \bar{m}(x) \equiv m^2 - (P_3-P_1)^2x(1-x)\:.
\end{displaymath}
Both $\bar{q}(x)$ and $\bar{m}(x)$ are small
throughout the range of integration of $x$.
Thus, the momentum integral is seen to be analogous to the triangle
diagram.  The procedure for separating $I_4$ into its regular 
and low-energy parts, $I_4=\bar{I}_4 +R_4$, 
is therefore identical to the prescription given for the triangle
diagram in the previous sub-section.

\section{Discussion}
\label{sec:Discussion}

It has been shown that each  basic one-loop integral  $I$ in the
two-heavy-particle sector can be decomposed into a low-energy part
$\bar{I}$  whose low-energy analytic structure coincides with that of
$I$,
and a regular part which consists of the terms in $I$ that have factors
of fractional powers of the large mass $M$.
The EFTDR  is now implemented by discarding $R$, which
amounts to removing specific  local terms in the low-energy domain.
The EFTDR so defined is a natural regularization.

The local terms in $\bar{I}$ can be ordered according to a power
counting
scheme in the low-energy scales generically denoted by $p$. For a given
one-loop  diagram the leading order term is of order $p^\nu$ with
$\nu=d-2 n_{\ell}-n_{\text{h}}+n_{\text{n}}$, 
where $n_{\ell}$ and $n_{\text{h}}$ are respectively the
numbers  of light and heavy propagators, and $n_{\text{n}}$ is the number of
factors
of small quantities in the numerator. This power counting is the
standard
one obtained in the $1/M$ expansion \cite{W2}.

The different non-analytic terms can be classified according to their
infrared behavior. In particular, some of them  are
more singular than the local terms. For example, the low-energy power
counting for the bubble diagram would be $\nu=2$, but Eq.~\eqref{IbarR}
shows that it is more singular than this.
If one takes $s-4 M^2$ to be of order $p^2$, the bubble diagram is of
order
$p$. In the case of the triangle, $\nu=0$, while $\bar{I}_{3 \text{A}}$
is more singular by one unit. As it is well known, these more singular
behaviors are associated with  the existence of the two-heavy-particle
threshold.

As first noticed in reference \cite{BL}, the fact that the low-energy
and regular parts have different behaviors in $M$
(the regular part being entirely of  fractional power of  M and the
low-energy part having no such terms), implies that
the symmetries respected by dimensional regularization
are also respected  by  each  part separately  and, therefore,
by EFTDR.

Finally, the  generalization to diagrams that involve insertions of
small
external  momenta can be carried out  quite easily. First, the
propagators
along each heavy particle line are separately combined using two sets of
Feynman parameters, and  a  parameter $w$ is then  used to combine the
two results. Next, and after combining the light particle propagators,
a second parameter $z$ is used to  combine this result with that of the
heavy propagators. The integration over the loop momentum is then
performed,
and subsequently the integration domain in $(w,z)$ is extended as in the case of the triangle
diagram, leaving the integration domains  of the remaining  Feynman
parameters unchanged.
A more  extensive presentation of this and other  points  in   this letter 
will be given elsewhere \cite{GLPS}.

\begin{ack}
  This work was supported by the Department of Energy through contract DE-AC05-84ER40150 (JLG),
  by the National Science Foundation through grants \#~PHY-9733343 (JLG and JS) and 
  \#~HRD-9633750 (DL and GP), by the Deutsche Forschungemeinschaft through project Le~1189/1-1 (DL), 
  and by sabbatical leave support from the Southeastern Universities Research Association (JLG). 
\end{ack}



\end{document}